\documentclass[12pt]{article}

\usepackage[english]{babel}
\usepackage{latexsym}
\usepackage{epsfig}

\begin{document}

\title{Decoherence in a spin--spin-bath model with environmental
self-interaction}

\author{L.~Tessieri and J.~Wilkie\\
{\it Department of Chemistry, Simon Fraser University,} \\
{\it Burnaby, British Columbia, Canada V5A 1S6} }

\date{11th September 2002}

\maketitle

\begin{abstract}
A low temperature model system consisting of a central spin coupled to a
spin-bath is studied to determine whether interaction among bath spins 
has an effect on central spin dynamics. In the absence of intra-environmental
coupling, decoherence of the central spin is fast and irreversible. Strong
intra-environmental interaction results in an effective decoupling of the
central spin from the bath and suppression of decoherence. Weaker
intra-environmental coupling reduces but does not eliminate decoherence.
We believe that similar behaviour will be observed in any system with a
self-interacting environment.
\end{abstract}


\section{Introduction}

Exact theories for the dynamics of open quantum systems require solution of
the Schr\"{o}dinger equation (or an equivalent formalism) for the full system
plus environment degrees of freedom. This is impossible to do analytically
except 
when the environment consists of a set of independent harmonic oscillators
(e.g., Caldeira-Leggett model~\cite{Cal83}, spin-boson model~\cite{Wei}) or
spins~\cite{Pro00}. Neglect of intra-environmental coupling is thus strongly
motivated by reasons of mathematical convenience. The purpose of the present
study is to determine whether this approximation is justified physically.
Our model consists of a small (i.e., highly quantum) central spin interacting
with a bath of self-interacting spins representing the degrees of freedom of
the environment. We shall sometimes refer to this central spin as the
subsystem.

Intuition suggests that coupling between environment degrees of freedom can
significantly affect the properties of the bath and consequently, through
subsystem-environment coupling, the dynamics of the central spin. Mutual 
interactions of the bath modes allow energy exchange without using the
subsystem as an intermediary. Consider an initial state of the
subsystem-environment which is perturbed away from equilibrium.
With intra-environ\-mental coupling the rapid initial flow of energy toward
a new equilibrium state will largely occur within the environment. Without
intra-environmental coupling the energy must flow through the subsystem,
leaving it strongly entangled with the bath and hence badly decohered.
[Note that the decoherence effects discussed in this manuscript differ
greatly from those which affect macroscopic subsystems~\cite{Braun}.]

Intra-environmental interactions may also alter more abstract properties of 
the bath. The classical dynamics of a bath of oscillators can change from
regular to chaotic when non-linear interactions are added~\cite{AA}. In the
quantum case, such coupling  modifies the statistical properties of the
energy levels and eigenstates of the bath. Specifically, the energy spectra
of quantum systems with chaotic classical counterpart exhibit level repulsion,
while systems with regular dynamics in the classical limit show a clustering
of the energy levels~\cite{Haa01}.
The structure of the bath eigenstates also changes when the dynamics
of the bath undergoes a transition from regular to chaotic. This effect
can be analysed by considering the form of the Wigner functions of
energy eigenstates: for chaotic systems the Wigner functions spread
more or less uniformly over the energetically available phase
space~\cite{Berry}, whereas in the case of regular systems they are more
lumpy. Moreover, these spectral signatures of chaos have dynamical
consequences~\cite{Wil97}.
It is therefore reasonable to expect that dissipation and decoherence
will assume different forms for coupled and uncoupled baths.

A qualitative understanding of the effects of bath self-interaction on
decoherence, while of obvious theoretical interest, might also have
important applications. Minimization of decoherence is essential for the
development and implementation of a number of new technologies
such as quantum computing~\cite{Lo998}, laser control of chemical
reactions~\cite{Bru98} and molecular electronics~\cite{Muj99}. For quantum
computing some proposed physical platforms, such as laser manipulated cold
ions in traps in near vacuum~\cite{Zoll} can very effectly minimise
decoherence. Whether such platforms can be scaled to the $10^{5}$ ions needed
to perform useful computations like factorisation of large integers is
unclear~\cite{Zoll}. Proposed solid state platforms such as single-electron
quantum dots embedded in a semiconductor~\cite{Bar} are readily
scaleable~\cite{Ber}, but decoherence is a serious obstacle. In principle
decoherence and dissipation can be reduced through judicious choice of
states for implementation of the qubit~\cite{Wha}. (A trivial example would
be to avoid states with electric-dipole allowed transitions.) Since the
solid state in principle provides enormous freedom over the choice of
qubit and matrix, further theoretical insight into the mechanisms of
decoherence might prove extremely useful in selecting optimal configurations. 

The effect of environmental self-interaction is almost certainly of
importance in the solid state. Unfortunately, as we noted above, the
analysis of the mechanisms of decoherence and dissipation in self-interacting
environments is a problem that defies exact analytical treatments.
Standard approaches, like the Feynman-Vernon influence functional
method~\cite{Fey63} or the Nakajima-Zwanzig projection technique~\cite{Nak58}
cannot be successfully applied unless restrictive hypotheses are made on the
nature of the bath and its coupling to the subsystem of interest. Thus,
for exact results we must rely on what we can calculate numerically.
Environments with few degrees of freedom should in principle provide much
of the qualitative information we seek. Such numerical studies also serve
a second purpose, namely to provide exact results which can be used as a
touchstone to test the reliability of the approximate analytical methods
which describe open systems (e.g., Redfield~\cite{Sua92},
SRA~\cite{Wil01} master equations and other approaches~\cite{Miller}).

In this paper we study numerically and without approximation the
dynamics of a spin-1/2 subsystem coupled to a bath of $N$ {\em interacting}
spin-1/2 modes. While spin--spin-bath models have been previously
studied~\cite{Pro00,Shao} intra-environmental coupling has been neglected.
Many different physical systems can be represented with such a model.
Examples include a spin chain, a set of magnetic impurities in a
solid~\cite{Bar} or an array of magnetic moments in a molecular crystal.
Alternatively, the system can be interpreted as a model of an atomic
impurity in a crystalline solid at low temperature. The impurity is
vibronically coupled to a bath consisting of the phonon modes of the
solid~\cite{Davies}. At low temperatures, the probability that the impurity
is in an electronic state other than the ground state is negligible. However,
the impurity can be prepared in a chosen (non-radiative) excited state and so
is effectively a two-level system. By the same token, in the low-temperature
limit the phonon modes reduce to oscillators with only two accessible levels
and can therefore be described as spin-1/2 modes (i.e., $a^{\dag} a
\rightarrow \sigma_{z}$, $a^{\dag} + a \rightarrow \sigma_{x}$, etc.).
Jahn-Teller interactions couple the central spin to the bath. Phonon-phonon
interactions inter-couple the bath spins. In our study we consider both
ferromagnetic and antiferromagnetic interactions.

We show that environmental self-interaction has an extremely important
effect on decoherence. Strong intra-environmental interactions effectively
decouple the subsystem from the environment, and even weak interactions
qualitatively change the manner in which phase information is lost. We see
no reason to suppose the effect is peculiar to the spin--spin-bath model,
and it should prove of importance in areas of physics where preservation of
quantum interference is paramount.

Section~\ref{moddef} explains our model in detail. In section~\ref{numapp}
we outline the numerical methods we employ to solve the subsystem dynamics.
Numerical results for subsystem dynamics are discussed in detail in
section~\ref{dynamics} for a wide range of intra-environmental couplings.
We summarise our findings in section~\ref{conclu} and discuss their
relevance for more general environments.

\section{The spin--spin-bath model}
\label{moddef}

In mathematical terms, our model is defined by the Hamiltonian
\begin{equation}
H = H_{0} + H_{B} + H_{I}
\label{ham}
\end{equation}
where
\begin{equation}
H_{0} = \frac{\hbar \omega_{0}}{2} \sigma_{z}^{(0)} + \beta \sigma_{x}^{(0)}
\label{hamzero}
\end{equation}
is the Hamiltonian of the central spin (denoted by the superscript $0$),
\begin{equation}
H_{B} = \sum_{i=1}^{N} \frac{\hbar \omega_{i}}{2} \sigma_{z}^{(i)}
+ \beta \sum_{i=1}^{N} \sigma_{x}^{(i)} +
\lambda \sum_{i=1}^{N-1} \sum_{j=i+1}^{N} \sigma_{x}^{(i)} \sigma_{x}^{(j)}
\label{batham}
\end{equation}
is the Hamiltonian of the bath, and
\begin{equation}
H_{I} = \lambda_{0} \sum_{i=1}^{N} \sigma_{x}^{(i)} \sigma_{x}^{(0)}
\label{intham}
\end{equation}
describes the interaction between the central spin and its environment.
For simplicity, in the rest of this paper we adopt a system of
units such that $\hbar = 1$.
Note that the terms proportional to $\beta$ and $\lambda$ in
Hamiltonians~(\ref{batham}) represent anharmonic
phonon-phonon interactions or non-linear spin interactions, depending
on which interpretation of the model one has in mind.

To complete the definition of the model, we have to specify the values
of the various parameters which appear in the Hamiltonian.
We assume that the frequencies of the bath spins are positive random
variables with the Debye probability density
\begin{equation}
p(\omega) = \left\{ \begin{array}{lcl}
3 \omega^{2}/\omega_{c}^{3} & \mbox{for} & 0 < \omega < \omega_{c} \\
0 & \mbox{for} & \omega_{c} < \omega \\
                    \end{array} \right. 
\label{debye}
\end{equation}
appropriate for the low energy acoustic modes of a crystal.
This is equivalent to saying that the bath has an Ohmic spectral density
with a sharp cut-off at $\omega = \omega_{c}$ which is not necessarily the 
Debye frequency itself. At the temperatures we consider, modes with large
frequency are unlikely to be populated and so it makes sense to choose a
smaller $\omega_{c}$ to reflect this fact. For the frequency of the central
spin, we arbitrarily chose the value $\omega_{0} = 0.8288 \omega_{c}$.
Note that the results obtained in this manuscript do not depend crucially
on the specific form~(\ref{debye}) of the frequency distribution. In fact,
we repeated our calculations with the box distribution
\begin{displaymath}
p(\omega) = \left\{ \begin{array}{lcl}
1/\omega_{c} & \mbox{for} & 0 < \omega < \omega_{c} \\
0 & \mbox{for} & \omega_{c} < \omega \\
                    \end{array} \right.
\end{displaymath}
and found the qualitative behaviour of the model unaltered by the change.

To simplify the form of mathematical expressions, we set $\omega_{c} = 1$,
$\lambda_{0} = 1$, and we varied the relative strength of the
intra-environmental interactions by letting the parameter $\lambda$ range
in the interval from $\lambda = 0$ (bath without internal spin-spin coupling)
to $\lambda = 10$ (strong bath self-interaction).
In addition to considering positive values of $\lambda$, which correspond
to antiferromagnetic interactions, we investigated the case of
ferromagnetic couplings, letting $\lambda$ assume negative values
in the interval $[- 10,0]$.

We set $\beta = 0.01$. An important consequence of the fact that $\beta$,
although small, is not zero, is that the Hamiltonian~(\ref{ham}) cannot be
reduced to block form. To understand this point, we observe that the
Hilbert space of the system~(\ref{ham}) is spanned by the vectors
\begin{equation}
\begin{array}{ccc}
|0\rangle & = & 
|0\rangle_{N} |0\rangle_{N-1} \ldots |0\rangle_{1} |0\rangle_{0}\\
|1\rangle & = &
|0\rangle_{N} |0\rangle_{N-1} \ldots |0\rangle_{1} |1\rangle_{0}\\
|2\rangle & = &
|0\rangle_{N} |0\rangle_{N-1} \ldots |1\rangle_{1} |0\rangle_{0}\\
& \vdots & \\
|2^{N+1}-1\rangle & = &
|1\rangle_{N} |1\rangle_{N-1} \ldots |1\rangle_{1} |1\rangle_{0}
\end{array}
\label{basis}
\end{equation}
where the symbols $|1\rangle_{i}$ and $|0\rangle_{i}$ denote the
`up' and `down' states of the $i-$th spin, i.e., the eigenstates
of the Pauli $z$-spin matrix $\sigma_{z}^{(i)}$ with eigenvalues
$+1$ and $-1$ respectively. Note that the basis states~(\ref{basis})
can be conveniently interpreted as binary representations of integer
numbers ranging from $0$ to $2^{N+1}-1$ if the vector
\begin{equation}
|k_{N}, k_{N-1}, \ldots, k_{0} \rangle =
|k_{N}\rangle_{N} |k_{N-1}\rangle_{N-1} \ldots |k_{0}\rangle_{0}
\label{vector}
\end{equation}
(with $k_{i} = 0,1$ for $i = 0, \ldots, N$) is associated to the
integer
\begin{equation}
k = k_{0} + 2 k_{1} + \ldots + 2^{N} k_{N} =
\sum_{n=0}^{N} 2^{n} k_{n} .
\label{number}
\end{equation}
For $\beta=0$ the Hamiltonian~(\ref{ham}) can be reduced to block
form by regrouping the basis vectors~(\ref{basis}) in two sets defined
by the condition that the states of each set have an even or odd number
of spins `up'. In fact, for $\beta = 0$ all terms of the
Hamiltonian~(\ref{ham}) have the effect of flipping either zero or two
spins at once, thereby leaving invariant the subspaces spanned
by the `even' and `odd' basis states.
The terms proportional to $\beta$, on the other hand, flip just one spin
and therefore connect the `even' and `odd' subspaces, making
the Hamiltonian~(\ref{ham})  irreducible.

\section{Numerical approach}
\label{numapp}

Assume that the bath is
initially in thermal equilibrium at temperature $T$ and that the central
spin is in the excited state $|1\rangle_{0}$. [In the case of an impurity
in a insulating solid such an initial state could be prepared using a fast
laser pulse with a frequency matching a transition of the impurity but lying
in the crystal's band gap.]
The initial density matrix of the whole system therefore has the product
form
\begin{equation}
\rho(0) = \rho_{0}(0) \otimes \rho_{B}(0) ,
\label{rhozero}
\end{equation}
with
\begin{eqnarray*}
\rho_{0}(0) = |1\rangle_{0} {}_{0}\langle1| & \mbox{and} &
\rho_{B}(0) = (1/Q) \exp (-H_{B}/kT)
\end{eqnarray*}
where $Q = \mbox{Tr}_{B} \left[ \exp (-H_{B}/kT) \right]$.
After introducing the notation
\begin{equation}
H_{B} |\phi_{n}^{(B)}\rangle = E_{n} |\phi_{n}^{(B)}\rangle
\label{batheig}
\end{equation}
for the $2^{N}$eigenvalues and eigenvectors of the bath Hamiltonian
and
\begin{equation}
|\psi_{n}(0)\rangle = |1\rangle_{0} \otimes |\phi_{n}^{(B)}\rangle
\label{initcon}
\end{equation}
for the corresponding initial conditions of the total system,
we can write the initial density matrix~(\ref{rhozero}) in the form
\begin{displaymath}
\rho(0) = \sum_{n=1}^{2^{N}}
|\psi_{n}(0)\rangle \frac{e^{-E_{n}/kT}}{Q} \langle \psi_{n}(0)|.
\end{displaymath}

To study the dynamics we have
determined the states
\begin{equation}
|\psi_{n}(t)\rangle = \exp \left( -iHt \right) |\psi_{n}(0)\rangle
\label{evolpsi}
\end{equation}
which evolve from~(\ref{initcon}).
Once states~(\ref{evolpsi}) are known the evolved density is constructed via
\begin{equation}
\rho(t) = \exp(-iHt) \rho(0) \exp(iHt)
= \sum_{n=1}^{2^{N}} |\psi_{n}(t)\rangle \frac{e^{-E_{n}/kT}}{Q}
\langle \psi_{n}(t)|,
\label{rho}
\end{equation}
and the reduced density of interest is
\begin{equation}
\rho_{0}(t) = \mbox{Tr}_{B} \left[ \rho(t) \right] = \sum_{n,m}
\langle \phi_{m}^{(B)} | \psi_{n}(t)\rangle \frac{e^{-E_{n}/kT}}{Q}
\langle \psi_{n}(t) | \phi_{m}^{(B)} \rangle.
\label{redrho}
\end{equation}

We used two
complementary techniques to compute the dynamics~(\ref{evolpsi}).
For small baths, $N \le 11$, we numerically diagonalised both the
bath Hamiltonian~(\ref{batham}) and the total Hamiltonian~(\ref{ham})
using standard Householder routines (see Ref.~\cite{Pre92}).
This method gives the exact reduced
density at all times and allows us to consider a range of temperature.
However, it cannot be used for baths composed of a large number $N$ of
spins since matrices
of size $2^{N} \times 2^{N}$ quickly exceed computer resources.
For large baths with $N \geq 12$ spins, we exploited the low-temperature
limit. For $kT \rightarrow 0$, high-energy eigenstates of
the bath are depleted and one can truncate the sum in Eq.~(\ref{rho}) to
the first $M$ terms, with $M \ll 2^{N}$, so that the density matrix takes
the form
\begin{equation}
\rho(t) \simeq \sum_{n=1}^{M} |\psi_{n}(t)\rangle \frac{e^{-E_{n}/kT}}{Q}
\langle \psi_{n}(t)| .
\label{approx}
\end{equation}
Thus a complete diagonalisation of the bath Hamiltonian
becomes unnecessary. We chose a bath temperature of $kT = 0.02$ for which
the number of terms $M$ needed in~(\ref{approx}) is 20.
We checked that neglected terms were irrelevant by evaluating the probability
that the bath be in an eigenstate of energy $E > E_{20}$. For baths with
$N \leq 11$ spins with $0 \leq \lambda \leq 10$ we obtained an upper bound
\begin{displaymath}
P(E_{B} > E_{20}) = 1 - \sum_{n=1}^{20} e^{-E_{n}/kT}/Q <  10^{-4},
\end{displaymath}
showing that bath eigenstates with $E>E_{20}$ are unpopulated.
As a further check, for baths composed of more than 11 spins we estimated
the ratio $R = p_{20}/p_{1} = \exp (E_{1}-E_{20})$.
For $N=14$,
$0 \leq \lambda \leq 10$ and $kT = 0.02$ the
probability ratio was less than
$10^{-6}$.

To determine the $M$ bath eigenstates of lowest energy, we used
ARPACK routines based on the Lanczos algorithm(see~\cite{arp}).
Evolved states~(\ref{evolpsi}) were calculated
using a Runge-Kutta algorithm of eighth order~\cite{dop}.
Neither the programs for the partial diagonalisation of the bath
Hamiltonian, nor the Runge-Kutta subroutine required that
the whole Hamiltonian matrix be stored in the computer memory, but only
that the matrix-vector product $H |\psi\rangle$ be defined.
This more efficient use of computer resources allowed us to consider baths
of up to 14 spins.

We calculated $H |\psi\rangle$ given an input state $|\psi\rangle$
by iterated calls to subroutines which multiplied by
$\sigma_{x}^{(i)}$ and $\sigma_{z}^{(i)}$.
Consider multiplication by $\sigma_{x}^{(i)}$ as an example.
With vectors~(\ref{basis}) as a basis for our Hilbert space
\begin{equation}
\sigma_{x}^{(i)} |\psi\rangle = \sum_{k=0}^{2^{N+1}-1}
\langle k | \psi \rangle \sigma_{x}^{(i)} | k \rangle,
\label{xmult}
\end{equation}
thus reducing our problem to that of finding an efficient way to
multiply the basis vectors~(\ref{basis}) by $\sigma_{x}^{(i)}$.
Since the matrix $\sigma_{x}^{(i)}$ has the effect of flipping the
i-th spin, one has
\begin{displaymath}
\sigma_{x}^{(i)} |k_{N},\ldots,k_{i},\ldots,k_{0}\rangle =
|k_{N},\ldots,\overline{k_{i}},\ldots,k_{0}\rangle
\end{displaymath}
where $\overline{k_{i}} = 1$ if $k_{i} = 0$ and $\overline{k_{i}} = 0$
if $k_{i} = 1$.
Multiplication by $\sigma_{x}^{(i)}$, therefore, replaces the
$k$-th component of $|\psi\rangle$ (where $k = k_{0} + \ldots + 2^{i} k_{i}
+ \ldots + 2^{N} k_{N}$) with the $k'$-th component (where $k' = 
k_{0} + \ldots + 2^{i} \overline{k_{i}} + \ldots + 2^{N} k_{N}$) and
vice-versa.
In binary representation, the numbers $k$ and $k'$ differ by a single
bit (the $i$-th bit) and one can therefore obtain $k'$ from $k$ using
Fortran intrinsic functions. Specifically, we used the XOR-function
(exclusive or) to flip the $i$-th bit of the $k$-th state.
Multiplication by $\sigma_{z}^{(i)}$ can be similarly implemented.

After determining the evolved density~(\ref{rho}), we traced out the bath
degrees of freedom to obtain the reduced density~(\ref{redrho}).
As indicators of quantum coherence, we chose the polarisation and entropy
of the central spin defined respectively as
\begin{equation}
\vec{P}(t) = \mbox{Tr} [\rho_{0}(t) \vec{\sigma}]
\label{polar}
\end{equation}
and
\begin{equation}
S_{0} (t) = - \mbox{Tr} \left[ \rho_{0}(t) \ln \rho_{0}(t) \right] =
- \frac{1}{2} \ln \left( \frac{1-P^{2}}{4} \right)
- \frac{P}{2} \ln \left( \frac{1+P}{1-P} \right) .
\label{entropy}
\end{equation}
Here $P =|\vec{P}|$ denotes the modulus of the polarisation vector $\vec{P}$.
Note that~(\ref{polar}) contains as much information as the reduced density
itself. In fact~(\ref{redrho}) can be expressed in terms of $\vec{P}$ via
\begin{displaymath}
\rho_{0}(t) = \frac{1}{2} \left( {\bf 1} + \vec{P}(t) \cdot
\vec{\sigma} \right) .
\end{displaymath}

\section{Dynamics of the central spin}
\label{dynamics}

Here we examine the effects of intra-environmental couplings on the dynamics
of the central spin. We consider the antiferromagnetic and the
ferromagnetic cases separately. High temperature results apply
only to the case where the spin-bath represents true physical spins.

\subsection{Antiferromagnetic interactions}

To evaluate the effect of antiferromagnetic interactions we calculated
the dynamics of the central spin for values of $\lambda$ ranging from
$\lambda = 0$ (uncoupled spins) to $\lambda = 10$ (strong coupling). As a
point of reference, note that in the absence of {\em subsystem-environment}
coupling $P_{z}(t)$ - initially one - undergoes periodic fluctuations to
slightly smaller values. The components $P_{x}(t)$ and $P_{y}(t)$ - initially
zero - oscillate about zero with the same period and similar small amplitude.

\begin{figure}[htp]
\begin{center}
\epsfig{file=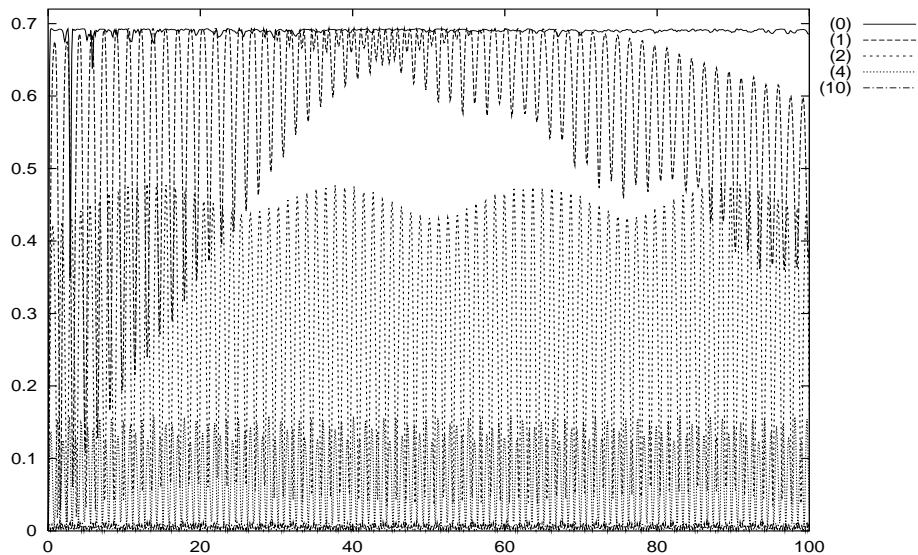,width=5in,height=3in}
\caption{$S_{0}(t)$ versus $\omega_{c}t$ (14 bath spins).}
\label{entro}
\end{center}
\end{figure}

\begin{figure}[htp]
\begin{center}
\epsfig{file=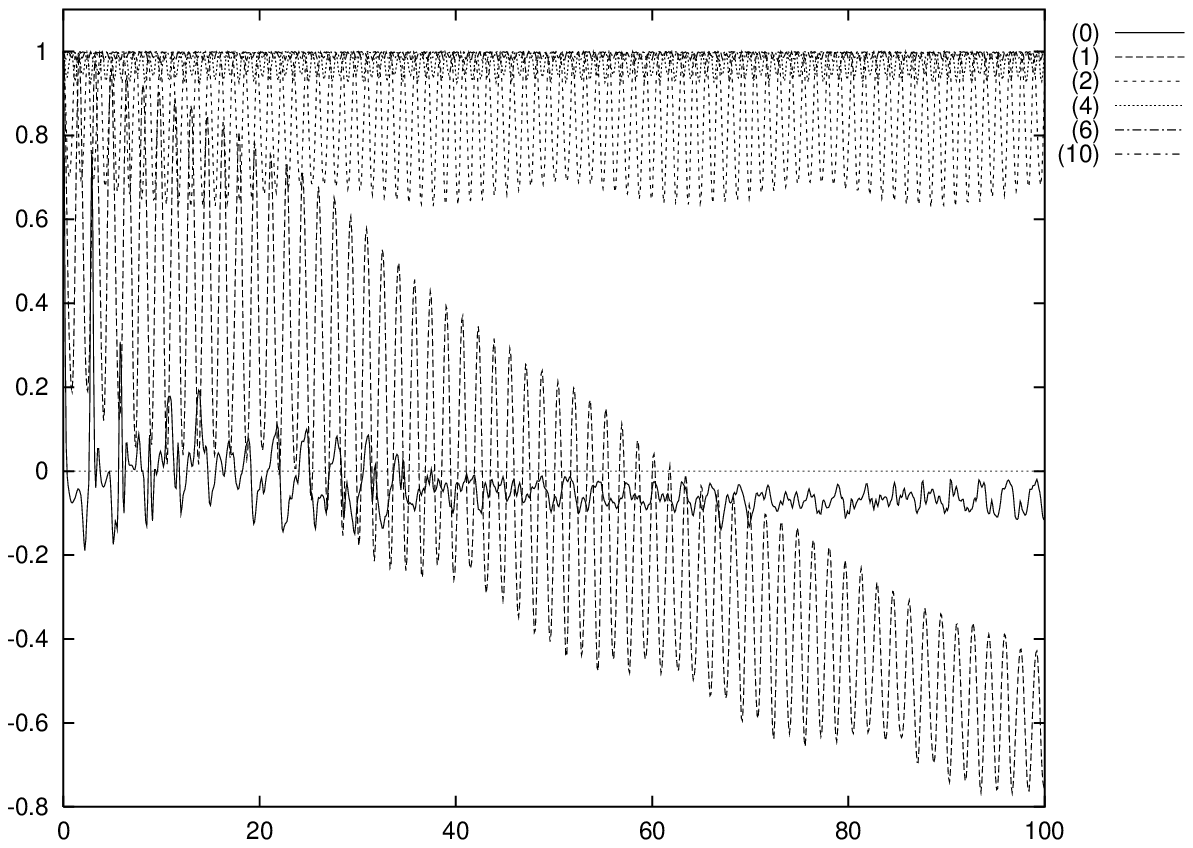,width=5in,height=3in}
\caption{$P_{z}(t)$ versus $\omega_{c}t$ (14 bath spins).}
\label{pz}
\end{center}
\end{figure}

\begin{figure}[htp]
\begin{center}
\epsfig{file=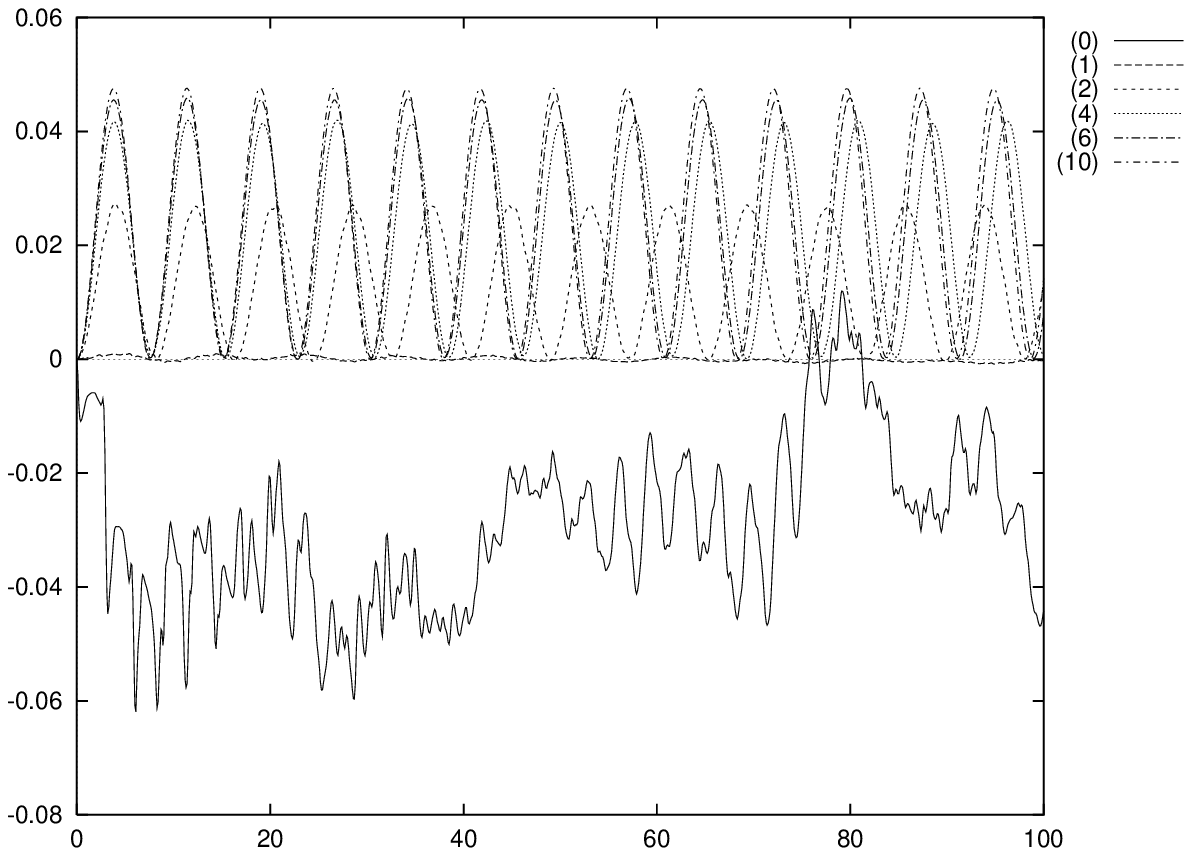,width=5in,height=3in}
\caption{$P_{x}(t)$ versus $\omega_{c}t$ (14 bath spins).}
\label{px}
\end{center}
\end{figure}

\begin{figure}[htp]
\begin{center}
\epsfig{file=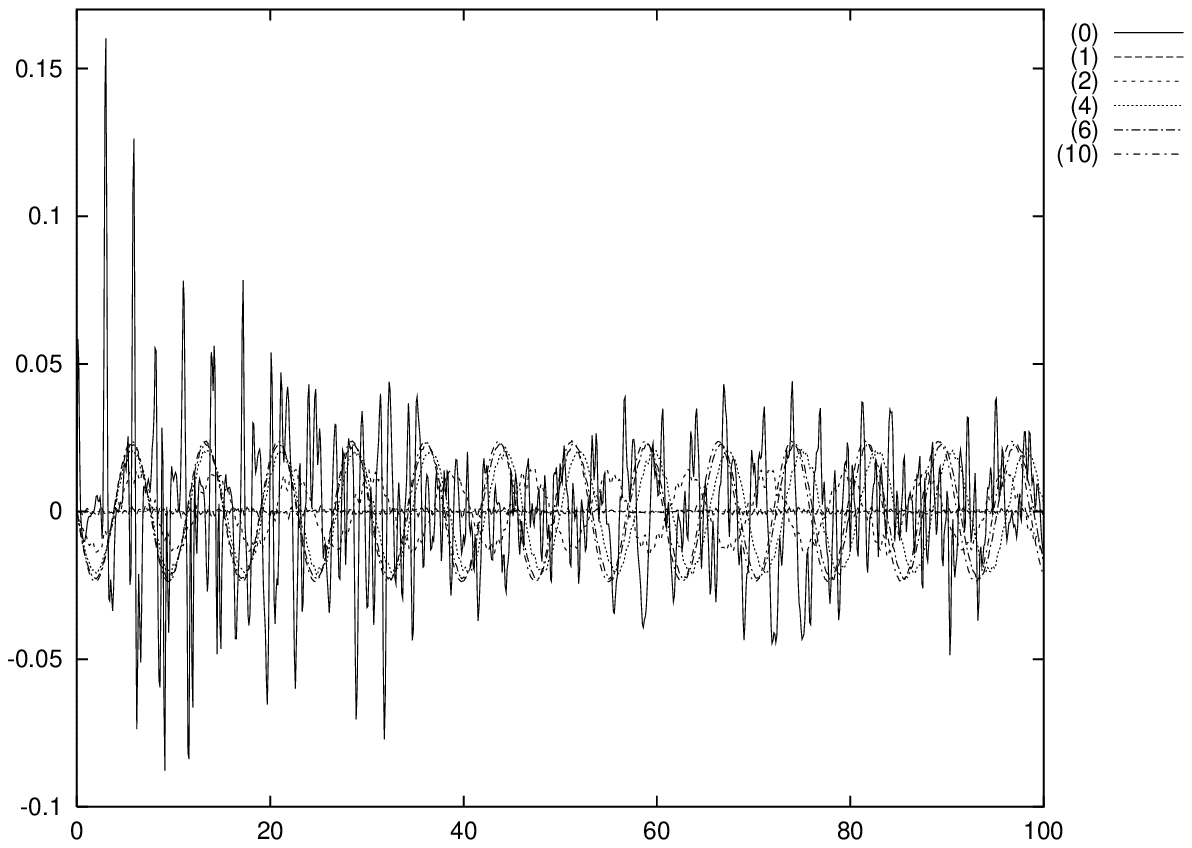,width=5in,height=3in}
\caption{$P_{y}(t)$ versus $\omega_{c}t$ (14 bath spins).}
\label{py}
\end{center}
\end{figure}

\begin{figure}[htp]
\begin{center}
\epsfig{file=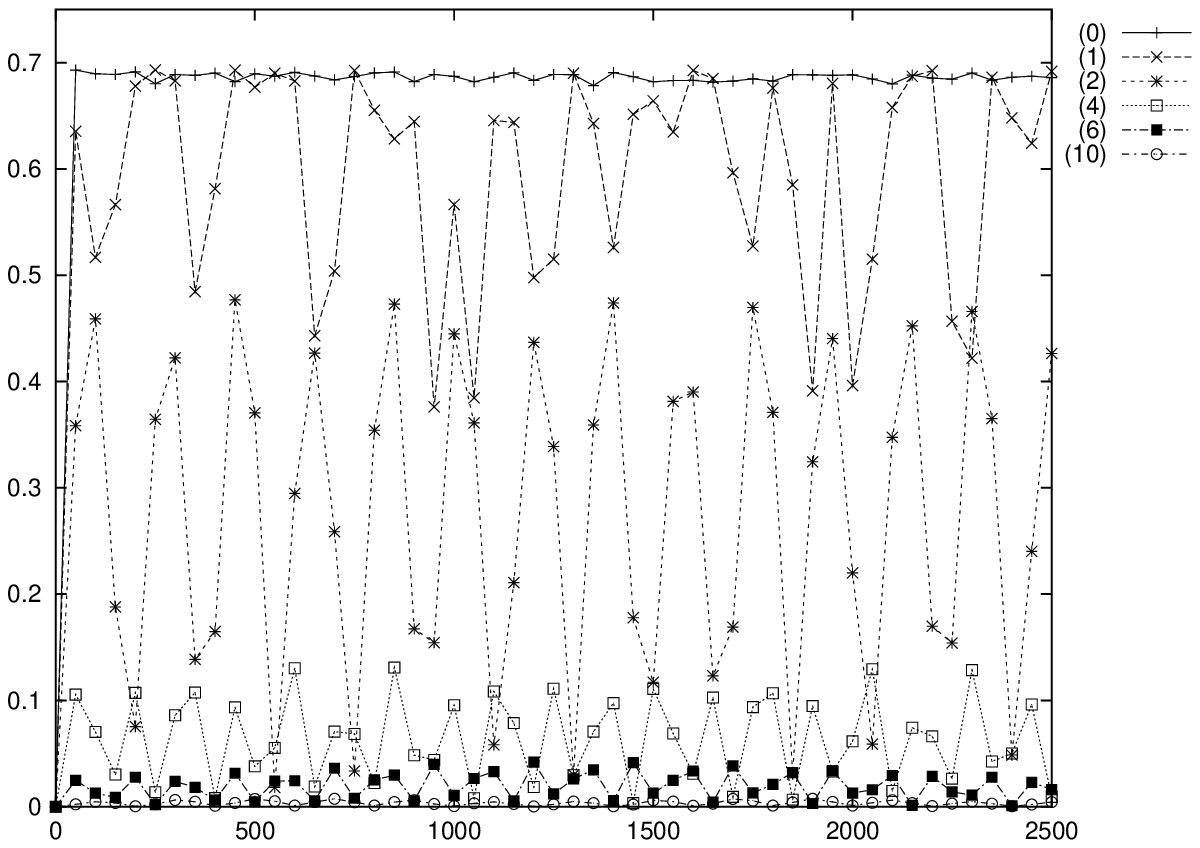,width=5in,height=3in}
\caption{$S_{0}(t)$ versus $\omega_{c}t$ (10 bath spins).}
\label{longent}
\end{center}
\end{figure}

The entropy~(\ref{entropy}) is shown in Fig.~\ref{entro}.
Figs.~\ref{pz},~\ref{px}, and~\ref{py} show components of the
polarisation~(\ref{polar}).
In the figures, each curve corresponds to a different value of the
intra-environmental coupling $\lambda$ (reported in parenthesis
to the right).
The thermal average was computed for a bath at temperature $kT = 0.02$.
The results shown are for $N=14$ but are also representative of smaller
baths with even numbers of spins (i.e., 8, 10 and 12).
The time scale considered is long enough for the system to attain
its asymptotic condition as can be seen from Fig.~\ref{longent}
which shows the entropy of the central spin on a much longer time scale
(data obtained for $N = 10$ spins).

When $\lambda = 0$ (i.e., bath spins non-interacting), the central spin
undergoes rapid decoherence, with the polarisation quickly falling to zero
and the entropy $S_{0}(t)$ simultaneously approaching the maximum value
$S_{0}^{max} = \ln(2) \simeq 0.693147 \ldots$.
As $\lambda$ increases, however, the entropy $S_{0}(t)$ tends to
progressively smaller asymptotic values and the components of the
polarisation vector approach the dynamics of a central spin evolving in
isolation. In other words, {\em the existence of a strong
interaction among bath spins suppresses environmental decoherence}.

\begin{figure}[htp]
\begin{center}
\epsfig{file=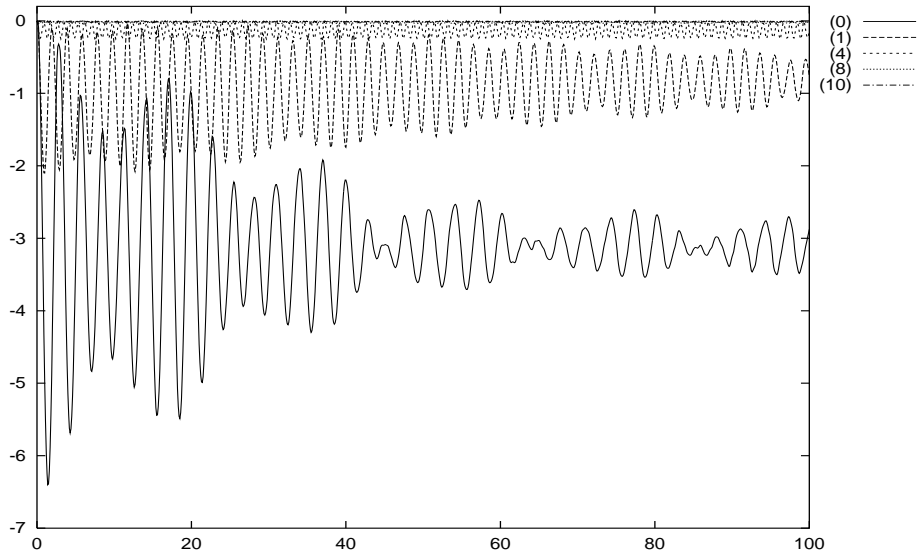,width=5in,height=3in}
\caption{Thermal average $\langle H_{I} \rangle$ of the interaction
Hamiltonian versus $\omega_{c}t$ (10 bath spins).}
\label{coupl}
\end{center}
\end{figure}

This apparently puzzling result is a straightforward consequence of the
fact that strong interactions between bath spins produce an almost
complete decoupling of the central spin from the bath.
This can be verified by considering the thermal average of
the interaction Hamiltonian~(\ref{intham}), defined as
\begin{equation}
\langle H_{I} (t) \rangle = \sum_{n=1}^{2^{N}}
\langle \psi_{n}(t) | H_{I} | \psi_{n}(t) \rangle e^{-E_{n}/kT}/Q .
\label{avint}
\end{equation}
Note that $\langle H_{I} (t) \rangle$, rather than $\lambda_{0}$, is the
physically relevant quantity determining the strength of the interaction.
In fact, the interaction may be small even if $\lambda_0$ is large.
The evolution of $\langle H_{I} (t) \rangle$ is displayed in
Fig.~\ref{coupl}, which shows that as $\lambda$ increases the
effective interaction of the central spin with the bath tends to zero
(data obtained for 10 spins).
Quantitatively, $\langle H_{I} \rangle$ averaged over the
time interval $[0:2500\omega_{c}^{-1}]$ decreases from
$\overline{\langle H_{I} \rangle} = -3.07$ for $\lambda=0$ to
$\overline{\langle H_{I} \rangle} = -0.01$ for $\lambda=10$.

There is a simple explanation for the observed behaviour of the average
interaction term~(\ref{avint}). Define the total bath spin
\begin{equation}
\vec{\Sigma} = \sum_{i=1}^{N} \vec{\sigma}^{(i)},
\label{macrospin}
\end{equation}
(i.e., $\Sigma_{x} = \sum_{i=1}^{N} \sigma_{x}^{(i)}$ etc.) and rewrite
the interaction Hamiltonian~(\ref{intham}) and bath Hamiltonian~(\ref{batham})
in the suggestive forms
\begin{displaymath}
H_{I} = \lambda_{0} \sigma_{x}^{(0)} \Sigma_{x}
\end{displaymath}
and
\begin{equation}
H_{B} = \frac{\lambda}{2} \left[ \Sigma_{x}^{2} - N {\bf 1} \right]
+ \beta \Sigma_{x} + \sum_{i=1}^{N} \frac{\omega_{i}}{2} \sigma_{z}^{(i)}.
\label{batham2}
\end{equation}
Eq.~(\ref{batham2}) shows that for $\lambda \gg \omega_{c}$ the
bath Hamiltonian takes the approximate form
\begin{equation}
H_{B} = \frac{\lambda}{2} \left[ \Sigma_{x}^{2} - N {\bf 1} \right]
+ O(\lambda^{0})
\label{batham3}
\end{equation}
which is essentially proportional to $\Sigma_{x}^{2}$.
Thus for large values of $\lambda$ the bath eigenstates~(\ref{batheig})
must be approximate eigenvectors of $\Sigma_{x}$ and those of lowest energy
must correspond to the eigenstates of $\Sigma_{x}$ with zero eigenvalue.
(That many such eigenstates exist is a consequence of the fact for every
value of the macroscopic variable $\Sigma_{x}$ there are many corresponding
microstates). Thus, for low temperature and large $\lambda$ the relevant
diagonal and off-diagonal matrix elements of the environment coupling
operator $\Sigma_x$ will be zero.

\begin{figure}[htp]
\begin{center}
\epsfig{file=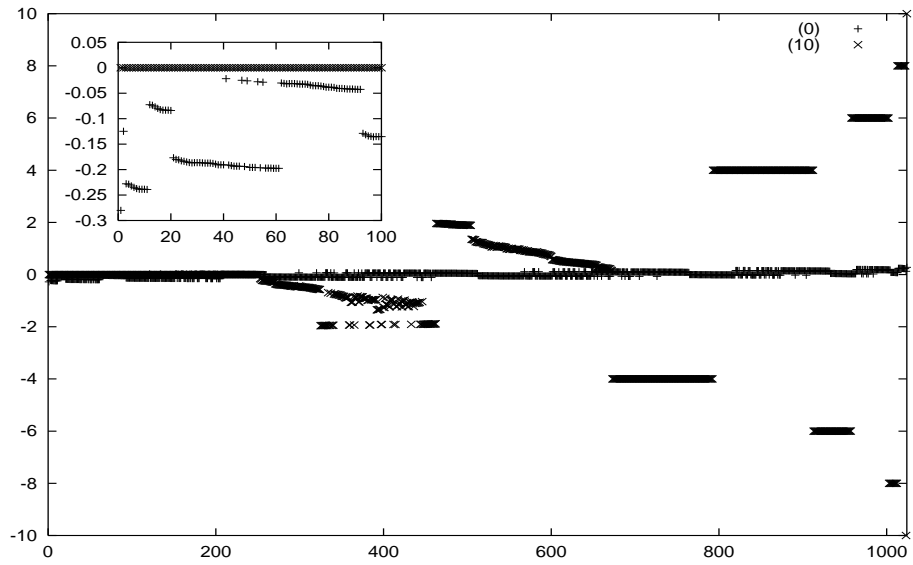,width=5in,height=3in}
\caption{Expectation value $\langle \phi_{n}^{(B)} | \Sigma_{x} |
\phi_{n}^{(B)} \rangle$ versus $n$ for $\lambda = 0$ and $\lambda = 10$.
The stacked inset shows on a larger scale the first 100 values.
The data were obtained for a 10-spin bath.}
\label{sxbath}
\end{center}
\end{figure}

These conclusions are confirmed by the numerical evaluation of the
expectation values of $\Sigma_{x}$ on the bath eigenstates~(\ref{batheig}).
Fig.~\ref{sxbath} reports $\langle \phi_{n}^{(B)}| \Sigma_{x}
|\phi_{n}^{(B)}\rangle$ as a function of the index $n$ which orders the
eigenstates $|\phi_{n}^{(B)}\rangle$ in ascending energy (data obtained
for $N = 10$ so $n$ runs from 1 to 1024).
Data are shown for $\lambda = 0$ and $\lambda = 10$.
For $\lambda = 0$ the expectation values of $\Sigma_{x}$ are small but
non-zero (this is more evident in the inset, which displays the data for
the 100 lowest energies). For $\lambda = 10$ the expectation value of
$\Sigma_{x}$ has a sort of step-like behaviour and attains relatively
large values for high-energy bath eigenstates. At low temperatures only
low-energy states matter and for these
$\langle \phi_{n}^{(B)}| \Sigma_{x} |\phi_{n}^{(B)}\rangle$
is orders of magnitude smaller than for the $\lambda = 0$ case (see inset
in Fig.~\ref{sxbath}). This effect can be quantified by computing the
quantity
\begin{equation}
\langle\Sigma_{x} \rangle = \frac{1}{20} \sum_{n=1}^{20}
\langle \phi_{n}^{(B)}| \Sigma_{x} |\phi_{n}^{(B)}\rangle .
\label{avsigma}
\end{equation}
One finds
\begin{equation}
\langle \Sigma_{x} \rangle =
\left\{ \begin{array}{lcc}
 -0.161951          & \mbox{for} & \lambda = 0 \\
 -6.0 \cdot 10^{-5} & \mbox{for} & \lambda = 10
        \end{array} \right.
\label{evensig}
\end{equation}
for $N = 10$ spins which is indeed small for strong coupling.

\subsubsection{High temperature limit}

\begin{figure}[htp]
\begin{center}
\epsfig{file=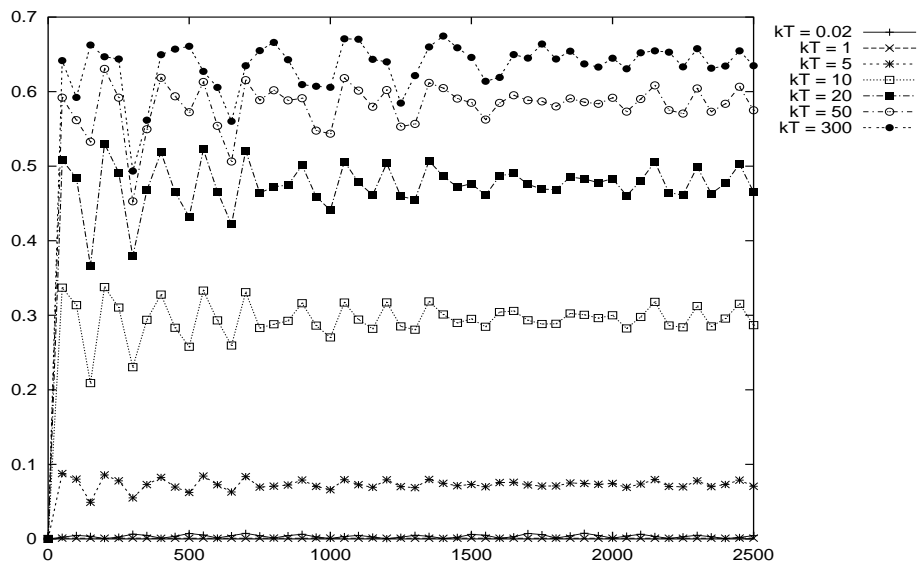,width=5in,height=3in}
\caption{$S_{0}$ versus $\omega_{c}t$ for $\lambda = 10$ for various
temperatures of the bath (10 bath spins).}
\label{tempent}
\end{center}
\end{figure}

\begin{figure}[htp]
\begin{center}
\epsfig{file=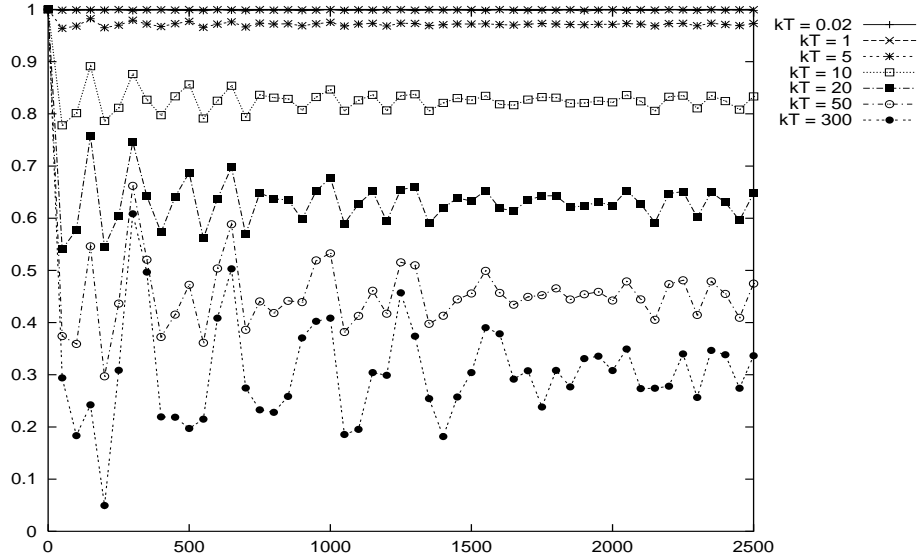,width=5in,height=3in}
\caption{$P_{z}$ versus $\omega_{c}t$ for $\lambda = 10$ for various
temperatures of the bath (10 bath spins).}
\label{temppz}
\end{center}
\end{figure}

Fig.~\ref{sxbath}, for $\lambda = 10$, shows that the expectation values
of $\Sigma_{x}$ are large for high-energy bath states. Populations in these
states are zero at low temperatures but increase with temperature.
The thermal average interaction~(\ref{avint}) will therefore also increase
with temperature, effectively coupling the central spin to its bath.
As a consequence, one can expect that the self-interacting bath will
behave more and more like an ordinary bath of uncoupled spins when
the temperature is raised. Fig.~\ref{tempent} shows the time behaviour of
the central spin entropy for a bath of $10$ coupled spins (with
$\lambda = 10$) at various temperatures.
Entropy increases monotonically with temperature, progressively approaching
the behaviour characteristic of the uncoupled bath. However, even at high
temperature ($kT = 300$) the entropy remains lower than for a bath
of uncoupled spins: this shows that even at high temperatures the
spin-spin coupling has a reductive effect on decoherence.
These features are confirmed in Fig.~\ref{temppz}.
Raising the temperature lowers the asymptotic value of $P_{z}(t)$ which
however never becomes zero.

In summary, at low temperatures the strong intra-environmental interactions
force the bath spins to align in an antiferromagnetic state of zero moment
and behave like a single giant spin decoupled from the central spin. As the
temperature is raised thermal fluctuations eliminate the spin alignment and
switch on the subsystem-environment coupling, making the self-interacting
bath behave more like an ordinary bath of uncoupled spins.

\subsubsection{Odd numbers of spins}

So far we have considered the behaviour of a central spin coupled to a
bath composed of an {\em even} number of spins.
To complete the discussion of antiferromagnetic interactions, we briefly
mention a last feature of the self-interacting spin bath - namely
modifications which appear for baths with an {\em odd} number of spins.

\begin{figure}[htp]
\begin{center}
\epsfig{file=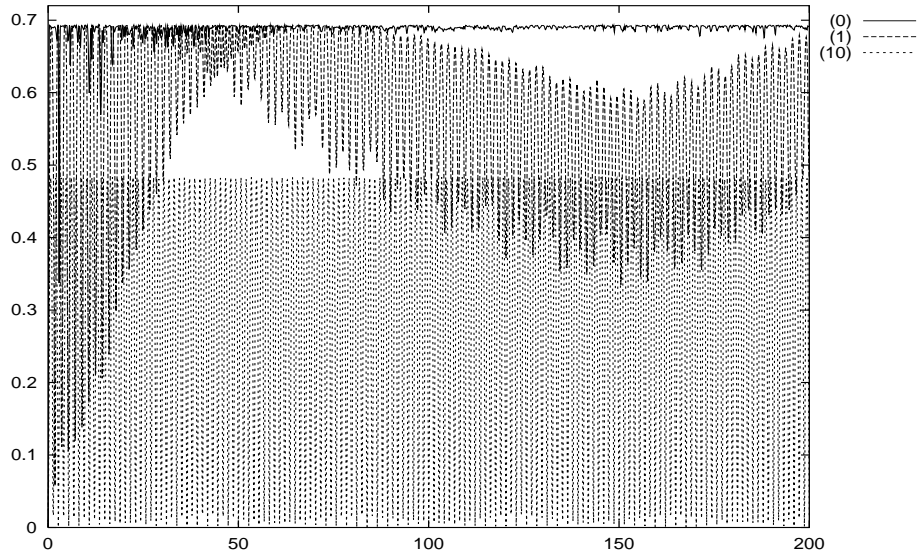,width=5in,height=3in}
\caption{$S_{0}$ versus $\omega_{c}t$ (bath composed of 11 spins).}
\label{oddentro}
\end{center}
\end{figure}

\begin{figure}[htp]
\begin{center}
\epsfig{file=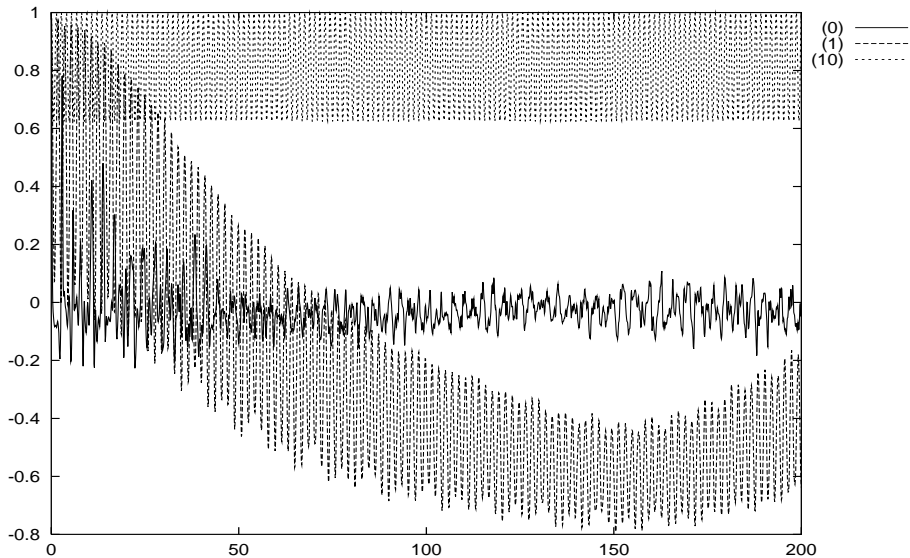,width=5in,height=3in}
\caption{$P_{z}$ versus $\omega_{c}t$ (bath composed of 11 spins).}
\label{oddpz}
\end{center}
\end{figure}

Figs.~\ref{oddentro} and~\ref{oddpz} show $S_{0}(t)$ and $P_z(t)$ for $N=11$.
Similar behaviour was obtained with baths of 5, 7, and 9 spins.
Only $\lambda=0,1$ and $10$ are shown because the curves for $\lambda$
greater than 1 overlap.
The data show that, regardless of whether the number of bath spins is even
or odd, $S_{0}(t)$ is a decreasing function of the intra-environmental
coupling. However, for $N$ odd, the average value of the entropy decreases
less than for $N$ even and oscillates more (compare Figs.~\ref{oddentro}
and~\ref{entro}). Polarisation behaves similarly.
As shown in Fig.~\ref{oddpz}, when the intra-environmental interactions are
strong, $P_{z}(t)$ oscillates around the value $P_{z}(t) \simeq 0.8$ for
baths with odd number of spins, whereas one has $P_{z}(t) \simeq 1$ for a
bath with an even number of spins.
Nor do the oscillations die on a longer time scale: in fact, numerical
calculations performed via the exact diagonalisation of the
Hamiltonian show oscillatory behaviour persisting on time scales an order
of magnitude longer than that in Figs.~\ref{oddentro} and~\ref{oddpz}.
Thus intra-environmental interactions have the effect of suppressing
decoherence regardless of how many spins compose the bath; however, this
effect is more marked if the bath is composed of an even, rather than an
odd, number of spins.

\begin{figure}[htp]
\begin{center}
\epsfig{file=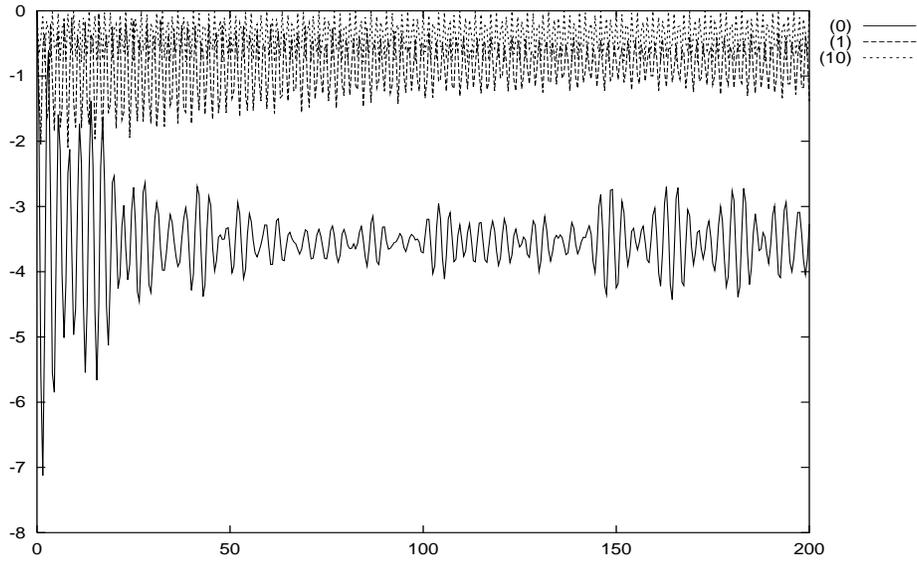,width=5in,height=3in}
\caption{Thermal average $\langle H_{I} \rangle$ of the interaction
Hamiltonian versus $\omega_{c}t$ (baths of 11 spins).}
\label{oddcoupl}
\end{center}
\end{figure}

The oscillations of entropy and polarisation are also displayed in the
strength of the interaction between the central spin and its bath.
Fig.~\ref{oddcoupl} shows the thermal average of the interaction
Hamiltonian~(\ref{intham}).
Comparing Fig.~\ref{oddcoupl} with Fig.~\ref{coupl}, it is evident that
the interaction between the central spin and bath is weaker for $N$ even
than $N$ odd.
Computing the parameter~(\ref{avsigma}) for a bath of 11 spins gives
\begin{equation}
\langle \Sigma_{x} \rangle =
\left\{ \begin{array}{lcc}
 -0.19180            & \mbox{for} & \lambda = 0 \\
 -4.97 \cdot 10^{-3} & \mbox{for} & \lambda = 10
        \end{array} \right. .
\label{oddsig}
\end{equation}
Comparing Eq.~(\ref{evensig}) with Eq.~(\ref{oddsig}) shows that
$\langle \Sigma_{x} \rangle$ is a decreasing function of $\lambda$
for baths of both even and odd numbers of spins but $\langle \Sigma_{x}
\rangle$ for $\lambda = 10$ is two orders of magnitude larger for the
11-spin bath than for the 10-spin bath. Table~\ref{table1} shows that as
$N$ increases $\langle \Sigma_{x} \rangle$ (for $\lambda = 10$) decreases.
The even-odd difference is thus a finite-size effect which should vanish
in the thermodynamic limit.
\begin{table}[htp]
\begin{center}
\begin{tabular}{|c|c|} \hline
Number of bath spins & $\langle \Sigma_{x} \rangle$ for $\lambda = 10$ \\
\hline
   7  &  -0.011147 \\
   9  &  -0.006724 \\
  11  &  -0.004967 \\
\hline
\end{tabular}
\caption{$\langle \Sigma_{x} \rangle |_{\lambda=10}$ as a function of $N$}
\label{table1}
\end{center}
\end{table}

\subsection{Ferromagnetic interactions}

\begin{figure}[htp]
\begin{center}
\epsfig{file=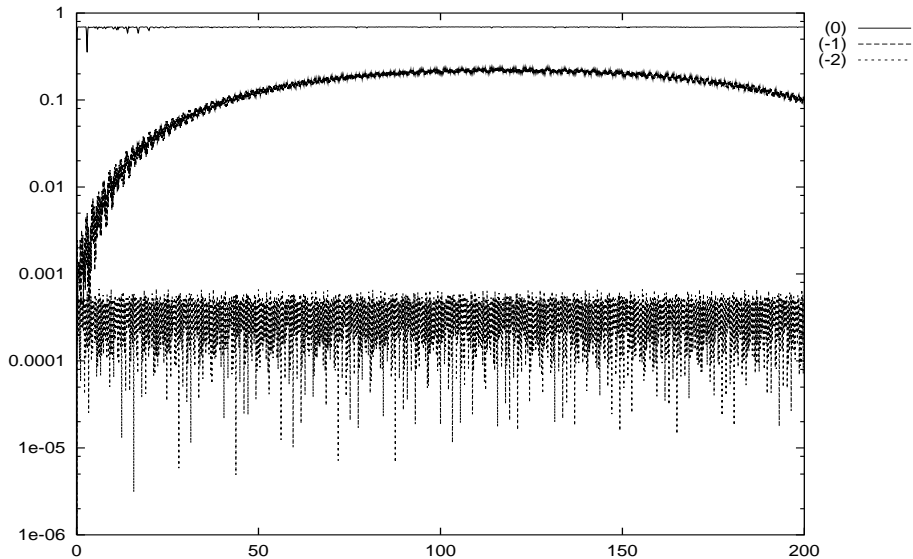,width=5in,height=3in}
\caption{$\ln \left( S_{0} \right)$ versus $\omega_{c}t$ (10 spin bath).}
\label{ferlogentro}
\end{center}
\end{figure}

The entropy for the ferromagnetic case is shown in Fig.~\ref{ferlogentro}.
The data were obtained for a bath of $10$ spins; we use a semi-logarithmic
scale to distinguish the curve for $\lambda = - 2$ from the $x$-axis.
We do not show data for values of $\lambda < -2$ 
because they overlap.
The behaviour of the entropy is similar in the ferromagnetic and
antiferromagnetic cases; the only significant difference is that in the
ferromagnetic case a suppression of the entropy can be achieved with
weaker intra-environmental interactions.

\begin{figure}[htp]
\begin{center}
\epsfig{file=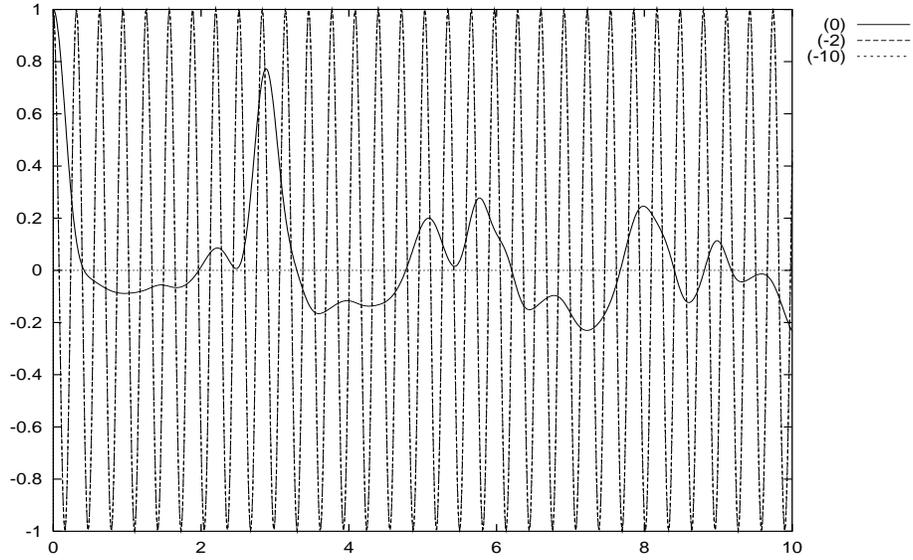,width=5in,height=3in}
\caption{$P_{z}$ versus $\omega_{c}t$ (10 spin bath).}
\label{ferpz}
\end{center}
\end{figure}

\begin{figure}[htp]
\begin{center}
\epsfig{file=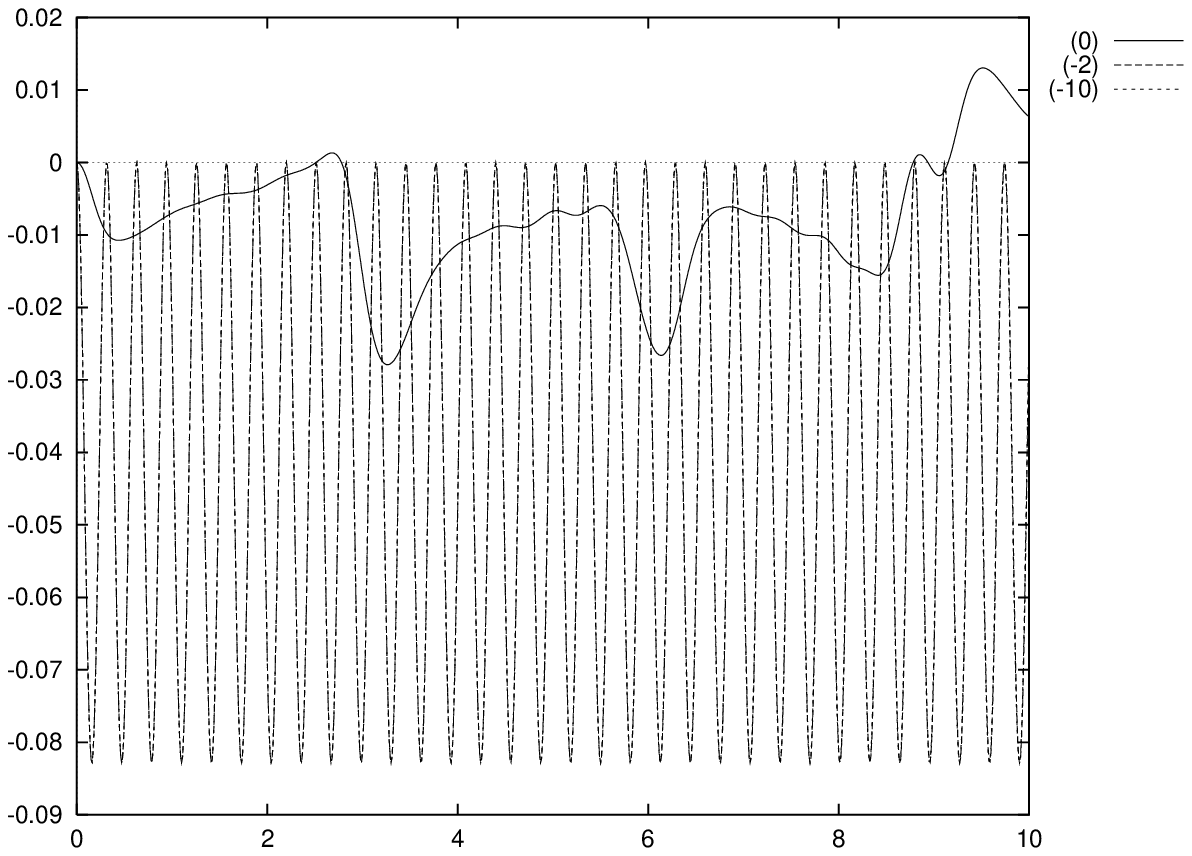,width=5in,height=3in}
\caption{$P_{x}$ versus $\omega_{c}t$ (10 spin bath).}
\label{ferpx}
\end{center}
\end{figure}

\begin{figure}[htp]
\begin{center}
\epsfig{file=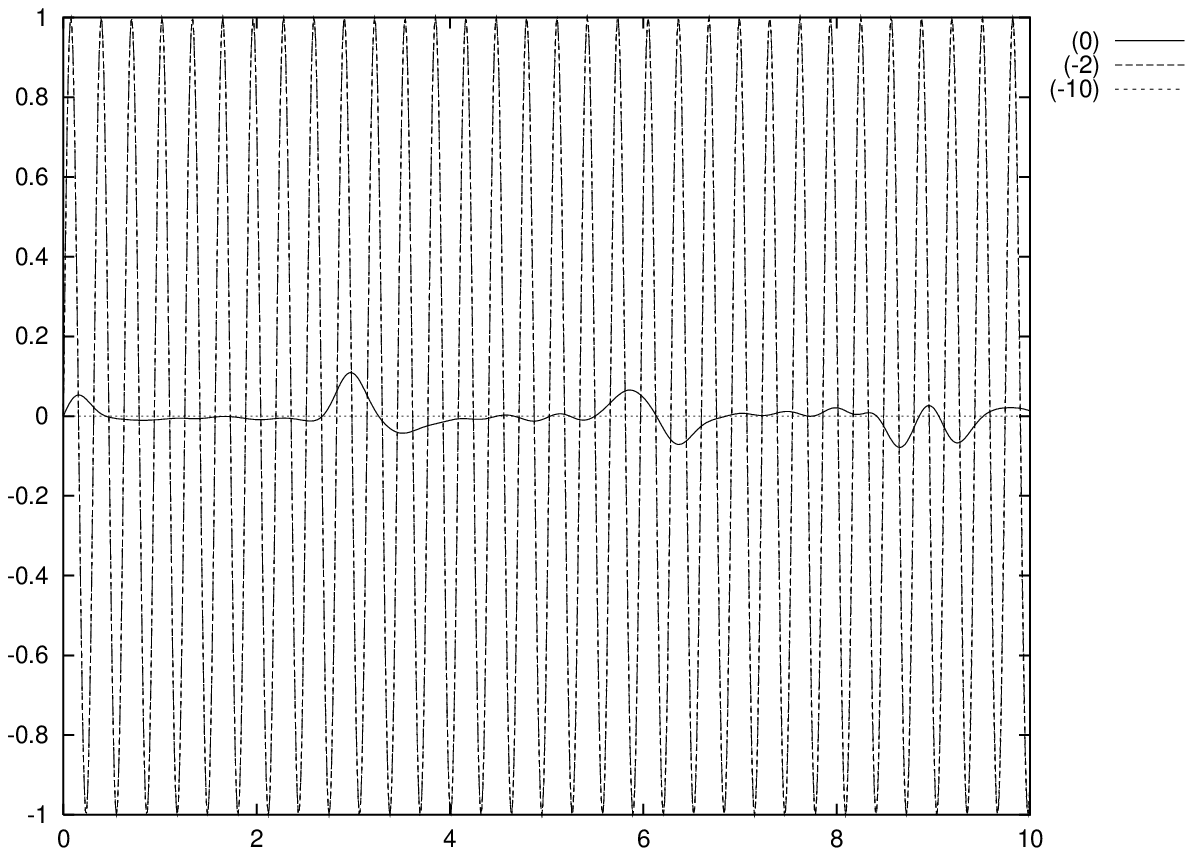,width=5in,height=3in}
\caption{$P_{y}$ versus $\omega_{c}t$ (10 spin bath).}
\label{ferpy}
\end{center}
\end{figure}

\begin{figure}[htp]
\begin{center}
\epsfig{file=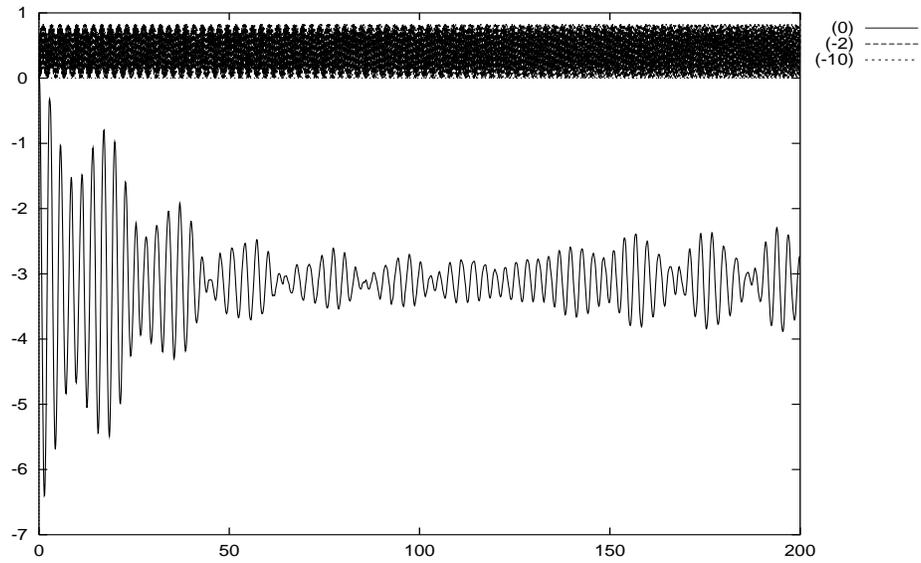,width=5in,height=3in}
\caption{$\langle H_{I} \rangle$ versus $\omega_{c}t$ (10 spin bath).}
\label{fercoupl}
\end{center}
\end{figure}

The most relevant difference between antiferromagnetic and ferromagnetic
baths emerges when one considers the behaviour of the polarisation
vector~(\ref{polar}) shown in Figs.~\ref{ferpz},~\ref{ferpx},
and~\ref{ferpy}. We have chosen a time scale appropriate to the fast
oscillations of the polarisation components (but long enough to be
representative of the long-time behaviour of $\vec{P}(t)$).
For $\lambda \leq -2$ the components of the polarisation vector assume
a strongly oscillatory behaviour. This is consistent with the suppression
of entropy since entropy depends only on the norm of the polarisation
vector which can stay close to unity even if individual components of
$\vec{P}$ oscillate in time.
Such polarisation dynamics, however, is very different from that
observed in the antiferromagnetic case where $P_{z}(t) \simeq 1$ and
$P_{x}(t) \simeq P_{y}(t) \simeq 0$ at all times. Term~(\ref{avint})
is not negligible for ferromagnetic interactions, as can be seen in
Fig.~\ref{fercoupl}, where we represent the thermal average of the
interaction Hamiltonian as a function of time.
For $\lambda \leq -2$, the interaction term $\langle H_{I}(t) \rangle$
becomes almost independent of the strength of the spin-spin coupling and
oscillates around a non-zero value.

\begin{figure}[htp]
\begin{center}
\epsfig{file=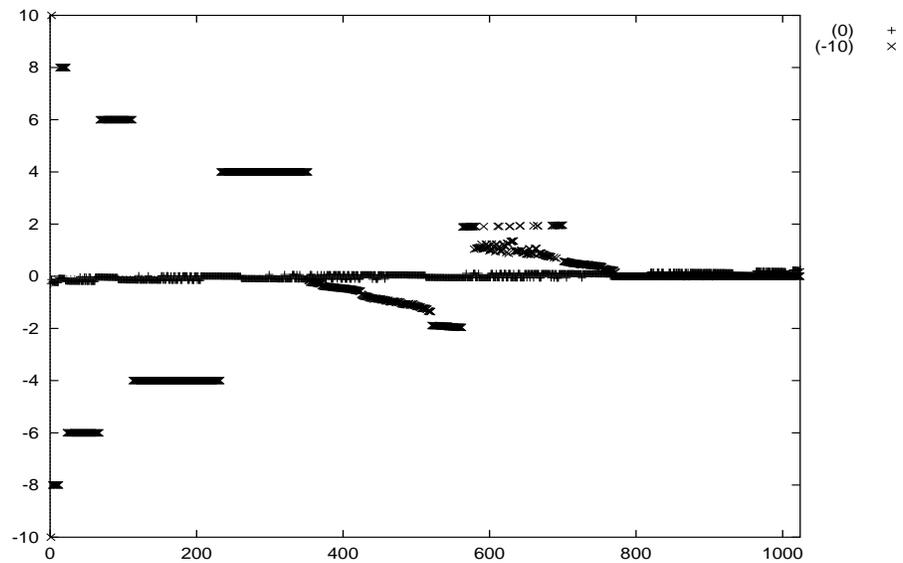,width=5in,height=3in}
\caption{Expectation value $\langle \phi_{n}^{(B)} | \Sigma_{x} |
\phi_{n}^{(B)} \rangle$ versus $n$ for $\lambda = 0$ and $\lambda = -10$
(10 spin bath).}
\label{fersxbath}
\end{center}
\end{figure}

This behaviour can be explained by writing the bath Hamiltonian in the
form~(\ref{batham2}) and by observing that in the limit of strong
ferromagnetic interaction
\begin{equation}
H_{B} = -\frac{|\lambda|}{2} \left[ \Sigma_{x}^{2} - N {\bf 1} \right]
+ O(\lambda^{0}).
\label{batham4}
\end{equation}
Note that the antiferromagnetic and ferromagnetic Hamiltonians~(\ref{batham3})
and~(\ref{batham4}) are identical but with opposite sign. The bath
eigenstates of lowest energy are eigenstates of $\Sigma_{x}$ in both cases,
but with
$\langle \phi_{n}^{B}| \Sigma_{x} | \phi_{n}^{B} \rangle \simeq 0$
in the antiferromagnetic case and
$|\langle \phi_{n}^{B}| \Sigma_{x} | \phi_{n}^{B} \rangle| \gg 1$
in the ferromagnetic case.
This is confirmed by Fig.~\ref{fersxbath}, which represents
$\langle \phi_{n}^{B}| \Sigma_{x} | \phi_{n}^{B} \rangle$
as a function of the index $n$.
Not surprisingly, the distribution of points for $\lambda = -10$ is
almost the reverse of the distribution for $\lambda = 10$ shown in
the corresponding Fig.~\ref{sxbath}. Consequently the coupling
term~(\ref{avint}) is significantly larger for a ferromagnetic bath than
it is for a antiferromagnetic one. Thus, diagonal matrix elements of the
coupling operator $\Sigma_{x}$ are large while off-diagonal matrix elements
are again zero.

It is important to note that the oscillations of the average interaction
Hamiltonian~(\ref{avint}) are synchronous with those of $P_{x}(t)$: more
precisely
\begin{displaymath}
\beta'(t) = \langle H_{I}(t) \rangle / P_{x}(t)
\end{displaymath}
is practically constant in time and, for $\lambda \leq -2$, it takes the
value
\begin{displaymath}
\beta' \simeq -10
\end{displaymath}
regardless of the strength of the spin-spin coupling. The value of $\beta'$
is easily explained. The matrix element
$\langle \phi_{1}^{B}| \Sigma_{x} | \phi_{1}^{B} \rangle$
corresponding to the lowest energy bath eigenstate is -10 while that of the
next lowest is 10. However, the energy gap between these states is .2 leading
to a population ratio $p_{2}/p_{1} \sim 10^{-5}$ at $kT=.02$. Hence, the
lowest energy state is the only one populated (in sharp contrast to the
antiferromagnetic case) and its $\Sigma_{x}$ eigenvalue is -10. This also
explains why decoherence is more easily suppressed in the ferromagnetic case.

Thus the central spin interacts with the bath through
\begin{equation}
H_{I}^{eff} = \beta' \sigma_{x}^{(0)},
\label{effham}
\end{equation}
and the evolution of the central spin is dictated by
Hamiltonian~(\ref{hamzero}) with a renormalised $\beta$ parameter:
\begin{equation}
\beta \rightarrow \tilde{\beta} = \beta + \beta' \simeq -10 .
\label{renbeta}
\end{equation}
Solving the Heisenberg equations for the renormalised central spin with
initial conditions $P_{z}(0) = 1$, $P_{x}(0) = P_{y}(0) = 0$ gives
\begin{equation}
\begin{array}{ccl}
P_{x} (t) & = & \displaystyle
\frac{2 \tilde{\beta} \omega_{0}}{\Omega^{2}} \left( 1 - \cos \Omega t
\right) \\
P_{y} (t) & = & \displaystyle
- \frac{2 \tilde{\beta}}{\Omega} \sin \Omega t \\
P_{z} (t) & = & \displaystyle
1 - \frac{4 {\tilde{\beta}}^{2}}{\Omega^{2}} \left( 1 - \cos \Omega t
\right)
\end{array}
\label{polev}
\end{equation}
where $\Omega = \sqrt{\omega_{0}^{2} + 4 {\tilde{\beta}}^{2}}$.
Comparing Eq.~(\ref{polev}) with the actual data shows that~(\ref{effham})
is correct. Figs.~\ref{ferpz},~\ref{ferpx} and \ref{ferpy} do not show
$\vec{P}(t)$ predicted by Eq.~(\ref{polev}) because of overlapping of the
various curves.

We therefore conclude that the only effect that a low temperature bath
with internal ferromagnetic interactions has on the central spin is a
renormalisation~(\ref{renbeta}) of the $\beta$ parameter in the
Hamiltonian~(\ref{hamzero}) (i.e., a Lamb shift).
In both the antiferromagnetic and the ferromagnetic cases, therefore,
when intra-environmental interactions are strong the dynamics of the
central spin is almost autonomous from that of the bath and is dictated
by a Hamiltonian of the form~(\ref{hamzero}).

As a last remark on the ferromagnetic case, we observe that the
dynamics of the central spin is unaffected by whether the bath is
composed of an even or odd number of spins.

\section{Summary and Conclusions}
\label{conclu}

In this work we considered a spin 1/2 subsystem coupled to a low-temperature
bath of interacting spin-1/2 modes. We focussed attention on the effects of
antiferromagnetic and ferromagnetic intra-environmental interactions on
decoherence. In both cases  strong intra-environmental interactions suppress
decoherence by making the dynamics of the central spin almost autonomous
from the bath itself.
More precisely, strong antiferromagnetic couplings among bath spins make
the average value of the subsystem-environment interaction Hamiltonian
vanish, thereby making the central spin evolve with its unperturbed and
uncoupled Hamiltonian. Strong ferromagnetic couplings among bath spins, on
the other hand, cause a Lamb shift of the subsystem Hamiltonian but
otherwise leave it to evolve in isolation.

These effects can be schematically explained by considering~(\ref{ham})
with $H_{I} = SB$ where $S$ and $B$ are subsystem and bath operators. Now,
the eigenstates of $H_{B}$ are also eigenstates of $B$ for strong
intra-environmental coupling in our model. Hence the off-diagonal matrix
elements of $B$ in this eigenbasis are zero and so~(\ref{ham}) cannot
couple bath eigenstates. Since the initial states of the full system are
proportional to bath eigenstates for initial conditions~(\ref{initcon})
it follows that the time evolved reduced density must be of the form
\begin{eqnarray}
\rho_{0}(t) = \sum_{n} \frac{e^{-E_{n}/kT}}{Q}~e^{-i(H_{0}+SB_{n})t}
\rho_{0}(0)e^{i(H_{0}+SB_{n})t}
\label{reff}
\end{eqnarray}
where $B_{n}$ is the eigenvalue of $B$ corresponding to eigenstate
$|\phi_{n}^{(B)}\rangle$ of $H_{B}$. In the antiferromagnetic case we
found that all relevant $B_{n}$ were zero for strong intra-environmental
coupling and hence the dynamics was free of decoherence. For the
ferromagnetic case we found that only the lowest energy state was populated
at low temperature and so only one term contributes to~(\ref{reff}) and
again the dynamics is coherent, but with a Lamb shift.

We expect similar effects to occur in more general baths of coupled
anharmonic oscillators. The key issue as we have seen is what happens to
the matrix elements of the spin-bath coupling Hamiltonian, in the eigenbasis
of the bath, when bath self-interactions are turned on. To suppress
decoherence the off-diagonal matrix elements must be small. Clearly in an
integrable bath (i.e., no self-interaction) some off-diagonal matrix
elements will be large due to selection rules. One would thus expect strong
decoherence for integrable environments. In the case of chaotic environments
it is known that the off-diagonal matrix elements are of order $h^{N-1}$
smaller than the diagonal matrix elements~\cite{Fein}, where $N$ is the
number of environmental modes. Since Planck's constant is small and $N$ is
very large, the off-diagonal matrix elements for a chaotic environment are
negligible.
Hence, we expect at most a Lamb shift of the subsystem for each eigenstate
of the bath and hence a reduced density like~(\ref{reff}). If in addition
the diagonal matrix elements (i.e. $B_n$) vary slowly with energy then at
low temperatures we should obtain a reduced density 
\begin{eqnarray}
\rho_{0}(t)=e^{-i(H_{0}+SB_{0})t}\rho_{0}(0)e^{i(H_{0}+SB_{0})t}
\end{eqnarray}
where $B_{0}$ is a representative low energy diagonal matrix element of
the coupling operator. Dynamics should therefore be coherent but (possibly)
Lamb-shifted for systems interacting with low temperature chaotic
environments.

Finally, returning to the specific case of an atomic impurity in a crystal
at low temperature, consider the limit of strong phonon-phonon coupling.
Since the Wigner functions of the bath eigenstates are nearly uniform over
the energetically available phase space~\cite{Berry}, energy is distributed
over an enormous number of phonon modes. At low temperature, displacements
from equilibrium of any phonon mode must therefore be small. Since coupling
of an impurity to a phonon is through its displacement
coordinate~\cite{Davies}, this coupling will also be small.
Thus, we expect diagonal and off-diagonal matrix elements of the Jahn-Teller
interaction to be small and decoherence to be minimal. Impurity-crystal
configurations which might have these attributes include noble gas
substitutional or interstitial impurities in diamond (which has strong
phonon-phonon interactions~\cite{Anas}).

The authors gratefully acknowledge the financial support of the Natural
Sciences and Engineering Research Council of Canada.

\end{document}